**Lorenz Number and Electronic Thermoelectric Figure of Merit: Thermodynamics and Direct DFT Calculations**

Yi Wang*, Jorge Paz Soldan Palma, Shun-Li Shang, Long-Qing Chen[1*], and Zi-Kui Liu
Department of Materials Science and Engineering, The Pennsylvania State University, University Park, PA 16802, USA

The Lorenz number ($L$) contained in the Wiedemann–Franz law represents the ratio of two kinetic parameters of electronic charge carriers: the electronic contribution to the thermal conductivity ($K_{el}$) and the electrical conductivity ($\sigma$), and can be expressed as $LT = K_{el}/\sigma$ where $T$ is temperature. We demonstrate that the Lorenz number simply equals to the ratio of two thermodynamic quantities: the electronic heat capacity ($c_{el}$) and the electrochemical capacitance ($c_N$) through $LT = c_{el}/c_N$, a purely thermodynamic quantity, and thus it can be calculated solely based on the electron density of states of a material. It is shown that our thermodynamic formulation for the Lorenz number leads to: i) the well-known Sommerfeld value $L = \pi^2/3(k_B/e)^2$ at low temperature limit; ii) the Drude value $L = (3/2)(k_B/e)^2$ at the high temperature limit with the free electron gas model, and iii) possible higher values than the Sommerfeld limit for certain semiconductors. It is also demonstrated that the purely electronic contribution to the thermoelectric figure-of-merit can be directly computed using high-throughput DFT calculations without resorting to the computationally more expensive Boltzmann transport theory to the electronic thermal conductivity and electrical conductivity.

*yuw3@psu.edu, lqc3@psu.edu

## I. Introduction

The Wiedemann–Franz law states that for solid electrical conductors, the ratio of the electronic contribution to the thermal conductivity ($\mathrm{K}_{el}$) to the electrical conductivity ($\boldsymbol{\sigma}$) is proportional to temperature ($T$), and the proportionality constant is the Lorenz number ($L$) [1,2]

$$\frac{\mathrm{K}_{el}}{\sigma} = LT \qquad \textbf{Eq. 1}$$

The law was established by Lorenz [1] based on the initial observation by Wiedemann and Franz [3] that "the conductivities for electricity and heat in metals are very closely related to each other, and are probably both functions of the same quantity". [4] One particular interest in the Lorenz number is concerning with the thermoelectric materials [5,6] for which there is a need to spilt the total thermal conductivity into the individual electronic and lattice contributions. In certain case, the Lorenz number has been assumed to be constant while it has been observed that its value can be reduced to half value of Sommerfeld limit. [7,8]

The first attempt to determine the theoretical value for the Lorenz number was made by Drude. [9] Using the free-electron model [10] and the classical high temperature limit in which the specific heat per charge carrier is constant, $L = 3/2(k_B/e)^2$, where $k_B$ is the Boltzmann's constant, $e$ is the elementary charge. On the other hand, using the Fermi-Dirac statistics and the low temperature limit in which the electronic heat capacity is linearly proportional to temperature, Sommerfeld [7] showed that $L = \pi^2/3(k_B/e)^2$.

While the earlier works assume the Lorenz number to be a constant universal to different materials, [4] modern theories [5,11,12] show that the Lorenz number can be temperature-dependent and varies with different materials, e.g., nondegenerate/degenerate semiconductors, multiband materials, or Fermi/non-Fermi liquids. Meanwhile, most published first-principles calculations of the Lorenz number based on the Boltzmann transport theory [13–15] assume constant relaxation time or the mean free path for electron scattering.[16]

In this work, we propose a simple theoretical argument to obtain the Lorenz number using well-defined thermodynamic quantities [17,18] including the electronic contribution to the heat capacity and the electrochemical capacitance. These thermodynamic quantities and thus the Lorenz number can be directly calculated based on the electron density of states from first-principles calculations.[19–23]

In Section II we theoretically demonstrate that the Lorenz number is simply the product of $k_B/e^2$ and the heat capacity per thermal carrier $c_{el}/n$ where $c_{el}$ is the specific heat capacity of thermal carriers and $n$ is the carrier density. The detailed computational formulism for Lorenz number is presented in Section III with only the electron density of states from DFT as the input. We shows in Section IV that the proposed theory recovers the Sommerfeld value as the low temperature limit, and in the free electron model, the Sommerfeld value is the upper limit and the Drude value is the lower limit; In Sec. V, we present calculated results based on density functional theory (DFT) [24] for selected pure metals and semiconductor $Si_{0.8}Ge_{0.2}$. We discuss the implications of the present theory to the efficient calculations of the electronic thermoelectric figure of merit in Section VI. Finally, the conclusions are presented in Section VII.

## II. Theory - revisit the Lorenz number

We now start the discussion from the thermal diffusivity of electrons ($\alpha_{el}$) [25] that can be used to relate the electronic contribution to the thermal conductivity and the electronic contribution to the specific heat by

$$K_{el} = c_{el}\alpha_{el} \qquad \textbf{Eq. 2}$$

In direct analogy to the thermal conductivity, we can write down a similar expression for the electric conductivity ($\sigma$) of electrons assuming that the electronic heat conduction and the electric conduction follows the same transport mechanism,

$$\sigma = c_N\alpha_{el} \qquad \textbf{Eq. 3}$$

where $c_N$ is the volumetric electrochemical capacitance. Therefore,

$$L = \frac{K_{el}}{T\sigma} = \frac{c_{el}\alpha_{el}}{Tc_N\alpha_{el}} = \frac{c_{el}}{Tc_N} \qquad \textbf{Eq. 4}$$

Thermodynamically, $c_{el}$ and $c_N$ are defined as

$$c_{el} = T\left(\frac{\partial s_{el}}{\partial T}\right)_{N_{el}} \qquad \textbf{Eq. 5}$$

and

$$c_N = \left(\frac{\partial q}{\partial \phi}\right)_T = e^2\left(\frac{\partial N_{el}}{\partial \mu}\right)_T \qquad \textbf{Eq. 6}$$

where $s_{el}$ is the volumetric entropy, $q = -eN_{el}$ is the charge densities of electrons, $N_{el}$ is the electron density, $\mu$ is the chemical potential of electrons, and $\phi = \mu/(-e)$ is the electrical potential. The derivation of Eq. 6 is supported by the Fick's first law [26]

$$J = -\alpha_{el}\nabla N_{el} \qquad \textbf{Eq. 7}$$

where $J$ represents the particle flux of electrons. Replacing the particle flux $J$ with the electrical current flux $j = -eJ$ followed by further utilizing $\nabla N_{el} = \left(\frac{\partial N_{el}}{\partial \mu}\right)_T \nabla\mu$ together with $\mu = -e\phi$, one gets

$$j = e^2\alpha_{el}\left(\frac{\partial N_{el}}{\partial \mu}\right)_T \nabla\phi = \alpha_{el}c_N\nabla\phi = \sigma\nabla\phi \qquad \textbf{Eq. 8}$$

III. Computational formulism based on electron density of states

Thermodynamically, one can obtain $(\partial N_{el}/\partial \mu)_T$ and $c_{el}$ of a material using only the electron density of states calculated from density functional theory (DFT). [24,27] The grand thermodynamic potential energy density $\Xi$ for an electron system at constant volume can be expressed as

$$d\Xi = -N_{el}d\mu - s_{el}dT \qquad \textbf{Eq. 9}$$

From the finite temperature DFT [27], the concentration of electrons $N_{el}$ in the system is determined by

$$N_{el} = \int_{-\infty}^{\infty} fD(\varepsilon)d\varepsilon \qquad \textbf{Eq. 10}$$

where $\varepsilon$ is the band energy, $D(\varepsilon)$ the volumetric electron density of states, and $f$ the Fermi-Dirac distribution

$$f = \frac{1}{\exp\left(\frac{\varepsilon - \mu}{k_B T}\right) + 1} \qquad \textbf{Eq. 11}$$

Therefore,

$$\left(\frac{\partial N_{el}}{\partial \mu}\right)_T = \frac{n}{k_B T} \qquad \textbf{Eq. 12}$$

where

$$n = \int_{-\infty}^{\infty} (1-f) f \, D(\varepsilon) d\varepsilon \qquad \textbf{Eq. 13}$$

which can be considered as the concentration of the effective mobile charges, or the effective concentration of electronic thermal carriers. Eq. 13 shows that only the electronic states near the Fermi level [27,28] contribute to the electric or thermal conduction. In other words, the electron system can be viewed as a system made up of effective mobile carriers, $n$, which makes the main contributions to the electronic heat conductivity, electronic heat capacity, and electronic electric conductivity. The concept of the mobile charge/electronic thermal carrier, $n$ in Eq. 13, is clearly supported by the understanding that the Lorenz number is related to the specific heat per mobile charge carrier ($C_{el}/n$) by a constant factor.

Following the previous works [17,29], $\boldsymbol{c_{el}}$ in Eq. 5 can be formulated into

$$\boldsymbol{C_{el}} = \frac{1}{k_B T^2}\left\{\int_{-\infty}^{\infty} (\varepsilon-\mu)^2 d\varepsilon (1-f) f D(\varepsilon) - \frac{1}{n}\left[\int_{-\infty}^{\infty} (\varepsilon-\mu) d\varepsilon (1-f) f D(\varepsilon)\right]^2\right\} \qquad \textbf{Eq. 14}$$

With Eq. 6 and Eq. 12, the Lorenz number in Eq. 4 becomes

$$L = \frac{k_B}{e^2}\frac{\boldsymbol{c_{el}}}{n} \qquad \textbf{Eq. 15}$$

IV. Model verification of the proposed theory

A. The degenerate limit

We first demonstrate that one can recover the Sommerfeld limit (refereed as the degenerate limit in literature) at low temperature from Eq. 14. Using Sommerfeld's low temperature expansion, [30,31] the specific heat $\boldsymbol{c_{el}}$ in Eq. 14 is reduced to

$$c_{el}(T \to 0) = \frac{\pi^2}{3} k_B^2 T D(\varepsilon_F) \qquad \textbf{Eq. 16}$$

where $\varepsilon_F$ is the Fermi energy, i.e., the value of $\mu$ at $T \to 0$, and $D(x)$ is volumetric electron density of states. The electrochemical capacitance in the low-temperature limit is given by

$$c_N(T \to 0) = e^2 \left(\frac{\partial N_{el}}{\partial \mu}\right)_T = e^2 \frac{n(T \to 0)}{k_B T} = e^2 D(\varepsilon_F) \qquad \textbf{Eq. 17}$$

Therefore at $T \to 0$, one gets

$$L(T \to 0) = \frac{c_{el}(T \to 0)}{T c_N(T \to 0)} = \frac{\frac{\pi^2}{3} k_B^2 T D(\varepsilon_F)}{T e^2 D(\varepsilon_F)} = \frac{\pi^2}{3} \left(\frac{k_B}{e}\right)^2 \qquad \textbf{Eq. 18}$$

which is the Sommerfeld value [7] for the Lorenz number which equals to $2.44 \times 10^{-8}$ W Ω $\mathrm{K}^{-2}$.

B. The non-degenerate limit

This section demonstrates that the present theory can recover the Drude value for the Lorenz number at the non-degenerate limit, *i.e.*, for ideal electron gas at high temperature. The Drude model referred by this work was from a textbook by Schcroft and Mermin [32] [Equation (1.54), Page 23]. We note that for ideal free electron gas at low temperature, one still obtains the Sommerfeld value [7] for the Lorenz number in Eq. 18 as the low temperature limit due to the Fermi distribution given in Eq. 11.

The Sommerfeld model was derived as the low temperature limit for the constant volume specific heat capacity with the Fermi-Dirac distribution while the Drude model [9,10] is based on the classical high temperature limit $(3/2)k_B$ for the constant volume specific heat capacity per free electron. We demonstrate that the present formulation in Eq. 15 also recovers the Drude model result. With the electron gas at the high temperature limit, $f \ll 1$ so that $(1-f)f \to f \to \exp\left(\frac{\mu-\varepsilon}{k_B T}\right)$ in Eq. 13 and Eq. 14. Furthermore, the electron density of states for free electron gas can be expressed as [33]

$$D(\varepsilon) = \begin{cases} 0, for\ \varepsilon < 0 \\ A\varepsilon^{1/2}, for\ \varepsilon \geq 0 \end{cases} \qquad \textbf{Eq. 19}$$

where $A$ in a constant. Accordingly, the Lorenz number in Eq. 15 becomes

$$\boldsymbol{L} = \frac{k_B}{e^2}\frac{1}{k_B T^2}\left\{\frac{\int_0^\infty \exp\left(\frac{\mu-\varepsilon}{k_B T}\right)\varepsilon^{5/2}d\varepsilon}{\int_0^\infty \exp\left(\frac{\mu-\varepsilon}{k_B T}\right)\varepsilon^{1/2}d\varepsilon} - \left[\frac{\int_0^\infty \exp\left(\frac{\mu-\varepsilon}{k_B T}\right)\varepsilon^{3/2}d\varepsilon}{\int_0^\infty \exp\left(\frac{\mu-\varepsilon}{k_B T}\right)\varepsilon^{1/2}d\varepsilon}\right]^2\right\}$$

$$= \left(\frac{k_B}{e}\right)^2\left\{\frac{\int_0^\infty \exp(-y^2)\, y^6 dy}{\int_0^\infty \exp(-y^2)\, y^2 dy} - \left[\frac{\int_0^\infty \exp(-y^2)\, y^4 dy}{\int_0^\infty \exp(-y^2)\, y^2 dy}\right]^2\right\} = \frac{3}{2}\left(\frac{k_B}{e}\right)^2 \qquad \textbf{Eq. 20}$$

which is exactly the same result by Drude. [10] Numerically the Drude value for the Lorenz number equals to $1.11 \times 10^{-8}$ W Ω K$^{-2}$ which is nearer to the value of $1.5 \times 10^{-8}$ W Ω K$^{-2}$ than the Sommerfeld value of $2.44 \times 10^{-8}$ W Ω K$^{-2}$ for non-degenerate semiconductors as given by Kim et al. [5]

Figure 1 shows the calculated Lorenz numbers for the 0 K Fermi energies at $\varepsilon_F = 8.617 \times 10^{-4}$, $8.617 \times 10^{-3}$, $8.617 \times 10^{-2}$, 0.2585, 0.4309, and 8.617 eV. Due to the proportionality between $L$ and $c_{el}/n$ as given in Eq. 16, the *y* axis on the left hand side is marked using the international unit while *y* axis on the right hand side is marked in $k_B$ in Figure 1. It demonstrates that the Drude value and the Sommerfeld value represent, respectively, the lower and upper limits of the Lorenz number for the free electron gas model.

### C. The scattering mechanisms under the non-degenerate limit

The Drude model referred by the previous subsection was limited to case given by Schcroft and Mermin.[32] The more general model[34] for the non-degenerate limit assumes the following three conditions:

i. $f = \exp\left(\frac{\mu-\varepsilon}{kT}\right)$;

ii. The density of electron states near the band edge is given by $g(\varepsilon) \propto \varepsilon^{1/2}$; and

iii. Together with the electron relaxation $\tau_m(\varepsilon) \propto \varepsilon^r$ where r, describes the energy dependence of the scattering time

These assumptions result in

$$L = (k_B/e)^2(r + 5/2) \qquad \textbf{Eq. 21}$$

For acoustic deformation potential scattering, i.e., r = -1/2, one finds $L = 2(k_B/e)^2$ which corresponds to the well-known case of a constant mean-free-path. For the case of a constant scattering time, $r = 0$, and one finds $L = \frac{5}{2}(k_B/e)^2$.

If an explanation has to be given on the type of the physical scattering process assumed in the present work at the non-degenerate limit, we find that the proposed approach for the by this work corresponds to $\tau_m(\varepsilon) \propto \varepsilon^{-1}$. This can be understood by means of "constant phase space volume approximation" by Heisenberg uncertainty principle. In fact, one observes that all the scattering processed can simply summarized by

$$v^n \tau = constant \qquad \textbf{Eq. 22}$$

where $v$ is the electron group velocity. By this condition, one gets*:* *n=0* means constant relaxation time approximation; and *ii) n=1* means constant mean free path approximation.

To explain the result for the extreme case at the high temperature limit with an earlier free electron gas model proposed by Drude,[32] an extension can be made to a case of *n*=2. Then the condition becomes $v^2\tau = \frac{mv}{m} v\tau = \frac{pl}{m} = constant$ where $m$ is the electron mass and l can be though as the mean free path. The product of pl reminds us of the phase space volume. Then the present work can be thought as constant phase space volume approximation. With the above "constant phase space volume approximation" explanation, it does result in *r = -1* if one treats $\varepsilon \propto 1/2m\, v^2$ at the high temperature limit.

D. Application to ideal semiconductor

We assume an ideal semiconductor model for which the electron density of states is made of the valence band, band gap, and conduction bands as follows

$$D(\varepsilon) = \begin{cases} A(-\varepsilon)^{1/2}, for\ \varepsilon < -E_g \\ 0, for\ -E_g < \varepsilon < 0 \\ A\varepsilon^{1/2}, for\ \varepsilon \geq 0 \end{cases} \qquad \textbf{Eq. 23}$$

where $E_g$ represents the band gap. Based on a typical value of 0.3 eV for $E_g$, Figure 2 shows the calculated Lorenz numbers for the 0 K Fermi energies at $\varepsilon_F = 8.617\times 10^{-4}$, $8.617\times 10^{-3}$, $8.617\times 10^{-2}$, 0.2585, 0.4309, and 8.617 eV. It is seen from Figure 2 that, at relative high temperatures, for the cases of at $\varepsilon_F = 8.617\times 10^{-4}$, $8.617\times 10^{-3}$, and $8.617\times 10^{-2}$ eV, the Lorenz number increases exponentially due to the electronic excitation from the valence band to the conduction band, so-called bipolar excitation, explaining the experimentally observed higher values [35] for the high Lorenz number than the Sommerfeld limit.

E. Mobile electronic thermal carriers

To understand the mobile charge/electronic thermal carrier *n* in Eq. 13, we plot the normalized mobile electronic thermal carriers ($n$/$N_c$) in Figure 3 and Figure 4 for the free electron gas model and ideal semiconductor model, respectively, where $N_c$ represents the density of charge carriers (electrons) counted at 0 K from $\varepsilon = 0$ to the Fermi energy. for the 0 K Fermi energies at $\varepsilon_F = 8.617\times 10^{-4}$, $8.617\times 10^{-3}$, $8.617\times 10^{-2}$, 0.2585, 0.4309, and 8.617 eV

$\varepsilon_F = 8.617\times 10^{-4}$ eV and $8.617\times 10^{-3}$ eV represent the cases of very low electron densities. With increasing temperature, the amount of the mobile thermal carriers increases rapidly and reaches a value equal to that of $N_c$ for $\varepsilon_F = 8.617\times 10^{-4}$ eV and $8.617\times 10^{-3}$ eV (Figure 3). For the case of ideal semiconductor plotted in Figure 4 at $\varepsilon_F = 8.617\times 10^{-4}$ eV and $8.617\times 10^{-3}$ eV , with increasing temperature, the amount of the mobile electronic thermal carriers can rapidly increase and become larger than $N_c$, due to the excitation of electrons from valence band with energies lower than -$E_g$. On the other hand, with higher $\varepsilon_F$ starting from 0.2585 eV, the number of electrons in the system is becoming enough for the system to start to approach to a metal as seen also by the Lorenz number plotted in Figure 1 and Figure 2.

## V. DFT calculations and Lorenz number

### A. Selected pure metals

For realistic applications of the present thermodynamic theory, we consider a number of classical pure metals: Ag, Al, Au, Cd, Cu, Li, Mo, Pb, Pt, Sn, W, and Zn in their ground states. DFT calculations are carried out employing the projector-augmented wave (PAW) method [36,37] as implemented in the Vienna *ab initio* simulation package (VASP, version 5.3). Spin-orbit interactions are not considered. The electron density of states is calculated with an energy cutoff of 400 eV and a Γ-centered *k*-mesh of ~80000 points/atom using the modified Becke-Johnson exchange potential in combination with PBE-correlation. [38] The crystal structures and lattice parameters from the Materials Project [39] have been used.

Figure 5 shows the calculated Lorenz numbers for the 12 metals. For the majority of the metals, the DFT calculated values are rather close to the Sommerfeld limit or the available experimental data from Kittel's textbook. [33] The exception is Sn, which is poor metal. The values for Pt and W show moderate deviations from the Sommerfeld limit, which could be ascribed to the fact the lattice contribution is appreciable for the case of W, amounting to 33% on top of that electronic contribution at room temperature as calculated by Chen et al. [40].

### B. Si0.8Ge0.2

For detailed application of the present thermodynamic theory for the Lorenz number, we consider the classical thermoelectric material $Si_{0.8}Ge_{0.2}$ as the prototype. DFT calculations are carried out employing PAW and the Spin-orbit interactions are not considered. The random structure of $Si_{0.8}Ge_{0.2}$ is mimicked by a 40-atom SQS (special quasi-random structures). [41] The lattice structure and atomic positions are first relaxed with an energy cutoff of 245.7 eV and a Γ-centered $k$-mesh of ~4000 points/atom under the local density approximation (LDA). [42] After that, the electron density of states is calculated with an energy cutoff of 400 eV and a Γ-centered $k$-mesh of ~80000 points/atom using the modified Becke-Johnson exchange potential in combination with LDA-correlation. [43,44] To have better description of the band gap ($E_g$), the CMBJ parameter has fixed to 1.104 based on the weighted average over the values of 1.09 and 1.16 for Si and Ge, respectively. With these CMBJ values, the calculations can reproduce the experimental band gaps for Si and Ge [45]. The therefore calculated band gap for $Si_{0.8}Ge_{0.2}$ is 1.08 eV in agreement with the value of 1.05 eV obtained by Ferhat et al. [46] from their least square fit to experimental data.

Figure 6 shows the calculated Lorenz numbers for $Si_{0.8}Ge_{0.2}$ at the $p$-type doping levels of $n_D$ =1×10$^{18}$, 1×10$^{19}$, 1×10$^{20}$, and 1×10$^{21}$ $e$/cm$^3$. These doping levels cover the metallic state ($n_D$ =1×10$^{21}$ e/cm$^3$) and the semiconductor states ($n_H$ =1×10$^{18}$, 1×10$^{19}$, and 1×10$^{20}$ $e$/cm$^3$). The doping has been implemented under the rigid band approximation [14] as schematically illustrated in Figure 7a by shifting the Fermi energy ($\varepsilon_F$) away from the valence band maximum ($\varepsilon_{VBM}$) to reach the desired carrier concentration

$$n_D = -\int_{\varepsilon_{VBM}}^{\varepsilon_F} D(\varepsilon)d\varepsilon \qquad \textbf{Eq. 24}$$

For the semiconductor doping levels of 1×10$^{18}$, 1×10$^{19}$, and 1×10$^{20}$ $e$ /cm$^3$, the Lorenz numbers first decreased from the Sommerfeld value of 2.44×10$^{-8}$ W Ω K$^{-2}$ from $T$ = 0 to ~1.5×10$^{-8}$ W Ω K$^{-2}$ with increasing temperature until the sharp increasing point (more evident for doping levels of 1×10$^{18}$ and 1×10$^{19}$) ascribed to the bipolar excitation. [35] In terms of $c_{el}/n$, this corresponds to a decrease from the Sommerfeld value of $\pi^2/3 = 3.29\ k_B$ into ~2 $k_B$ from 0 K to the intermediate temperature.

Both $n$ in Eq. 13 and $c_{el}$ in Eq. 14 are (i) linear with respect $T$ at low temperature; (ii) almost constant at intermediate temperature; and (iii) increase exponentially with temperature (stronger than $n$) $T$ at high temperature, accounting for the bipolar excitation [35] (i.e. the excitation of electrons from valence bands to conduction bands). The ratio of the mobile charge carrier concentration to the nominal charge carrier concentration due to doping, $n/n_D$, are plotted in Figure 7b and the specific heats per nominal charge carrier, $c_{el}/n_D$, are plotted in Figure 7c. At low temperature, both $n/n_D$ and $c_{el}/n_D$ show linearities vs $T$, obeying the metallic behavior and therefore validating the Sommerfeld model. In-depth analysis can be made by decomposing the concentration of the mobile charge carriers in Eq. 13 into

$$n = n_h + n_e \qquad \textbf{Eq. 25}$$

where

$$n_h = \int_{-\infty}^{\varepsilon_{VBM}} (1-f)f\, D(\varepsilon) d\varepsilon \qquad \textbf{Eq. 26}$$

$$n_e = \int_{\varepsilon_{CBM}}^{\infty} (1-f)f\, D(\varepsilon) d\varepsilon \qquad \textbf{Eq. 27}$$

Knowing that when $T \rightarrow \infty$, the value $\mu$ in Eq. 12 by means of Eq. 11 is located in the range between $\varepsilon_{VBM}$ and $\varepsilon_{CBM}$. In such case, $f \rightarrow 1$ and 1-$f$ → exp $\left[\frac{\mu-\varepsilon}{k_B T}\right]$ in Eq. 24 while 1-$f$ → 1 and $f$ → exp $\left[-\frac{\mu-\varepsilon}{k_B T}\right]$ in Eq. 25. As a result, one gets the following expressions for the hole and electron charge carrier concentration [33] in the integral form

$$n_h(T \rightarrow \infty) = \int_{-\infty}^{\varepsilon_{VBM}} \exp\left[\frac{\varepsilon - \mu}{k_B T}\right] D(\varepsilon) d\varepsilon \qquad \textbf{Eq. 28}$$

$$n_e(T \rightarrow \infty) = \int_{\varepsilon_{CBM}}^{\infty} \exp\left[\frac{\mu - \varepsilon}{k_B T}\right] D(\varepsilon) d\varepsilon \qquad \textbf{Eq. 29}$$

where $\varepsilon_{CBM}$ is the conduction band minimum. The temperature dependences of $n_h$ and $n_e$ calculated respectively using Eq. 24 and Eq. 25 are illustrated in Figure 8 using the case of $n_D$ =1×10$^{18}$ $e$/cm$^3$. It is evident that $n_h$ has the meaning of hole concentration and $n_e$ has the meaning of electron concentration.

At the intermediate temperature range (depending on the doping levels), $n_h \cong n_D$ for p-type doping. This explains why $n/n_D$ becomes constant at the intermediate temperature range. It should note that $c_{el}/n_D$ is close to a constant for the low doping levels of 1×10$^{18}$, 1×10$^{19}$, and 1×10$^{20}$ at the intermediate temperature range.

At high temperatures where $n$ starts to increase exponentially from zero (depending on the doping levels), bipolar excitation is initiated, i.e., the electrons start to occupy the conduction bands, shown as the increase in both $n/n_D$ and $c_{el}/n_D$ (more evident for doping levels of 1×10$^{18}$ and 1×10$^{19}$).

# VI. Thermoelectric materials

## A. Relation between Lorenz number and Seebeck coefficients

It has been of interested studying the dependence of the Lorenz number on the Seebeck coefficient ($S_e$) . Kim et al. empirically found out the Lorenz number $L = 1.5 + \exp\left(-\frac{|S_e|}{115}\right)$ which are claimed to be within 5% for single parabolic band/acoustic phonon scattering assumption and within 20% for PbSe, PbS, PbTe, and Si0.8Ge0.2 In the previous work, [47] it has been demonstrated that the Seebeck coefficient is a thermodynamic quantity and can be calculated solely based on the electron density of states

$$S_e = -\frac{1}{enT}\int_{-\infty}^{\infty}(\varepsilon-\mu)(1-f)fD(\varepsilon)d\varepsilon \qquad \textbf{Eq. 30}$$

In fact, with the above $S_e$, the $C_{el}$ in Eq. 14 can be also reformulated as

$$C_{el} = C_{\mu} - \frac{ne^2}{k_B}S_e^2 \qquad \textbf{Eq. 31}$$

where $C_{\mu}$ is a kind of heat capacity under the condition of constant chemical potential used in the BoltzTrap2 code. [48]

$$C_{\mu} = \frac{1}{k_B T^2}\int_{-\infty}^{\infty}(\varepsilon-\mu)^2 d\varepsilon(1-f)fD(\varepsilon) \qquad \textbf{Eq. 32}$$

Then the Lorenz number in Eq. 15 becomes

$$L = \frac{k_B}{e^2}\frac{C_{\mu}}{n} - S_e^2 \qquad \textbf{Eq. 33}$$

Figure 9 illustrates the decompositions of the Lorenz numbers into the $C_{\mu}/n$ and $S_e^2$ contribution to for *p*-type $Si_{0.8}Ge_{0.2}$ at 300 K as functions of Seebeck coefficient. Note that this calculation has been performed by changing the doping level since the temperature is fixed. It shows that the calculation reproduced well the observed tendencies for typical thermoelectric materials [5] on the dependence of the Lorenz number on Seebeck coefficient.

## B. Electronic contribution to figure of merit of thermoelectric materials

The knowledge of the Lorenz number [4] is essential in the field of thermoelectrics dealing with the energy conversion between heat and electrics. [16] For example, to study the physical reason behind the performance of a thermoelectric material, it is often needed to separate the total thermal conductivity ($K$) into the electronic contribution and the lattice contribution ($K_{lat}$) throght $K = K_{el} + K_{lat} = \sigma LT + K_{lat}$. The performance of a thermoelectric material is quantified by the figure of merit [49,50]

$$zT = \frac{\sigma S_e^2}{K_{el} + K_{lat}} T \qquad \textbf{Eq. 34}$$

where $K_{lat}$ is the lattice contribution of the thermal conductivity, and $S_e$ is the Seebeck coefficient. Using the Lorenz number, the figure of merit can be rewritten as

$$zT = \frac{zT_{el}}{1 + K_{lat}/K_{el}} \qquad \textbf{Eq. 35}$$

where

$$zT_{el} = \frac{S_e^2}{L} \qquad \textbf{Eq. 36}$$

is called the purely electronic thermoelectric figure of merit. [16]

Consequently, $zT_{el}$ can be evaluated without calculating the electronic contributions to the thermal conductivity and electrical conductivity that require the kinetic quantities in the Boltzmann transport theory. [13–15] The calculated values of $zT_{el}$ for $Si_{0.8}Ge_{0.2}$ at the *p*-type doping levels of $n_D$ =1×10$^{18}$, 1×10$^{19}$, 1×10$^{20}$, and 1×10$^{21}$ $e$/cm$^3$ are plotted in Figure 10. It shows that for the doping levels of $n_D$ =1×10$^{18}$, 1×10$^{19}$, and 1×10$^{20}$ $e$/cm$^3$, $zT_{el}$ can be substantially larger than 1. Meanwhile by experiment, it was observed that for *p*-type $Si_{0.8}Ge_{0.2}$ the $zT$ value was mostly around 0.6. [51–53] This tells us that for most of the existing $Si_{0.8}Ge_{0.2}$, the lattice thermal conductivity must be very high when being compared with the electronic thermal conductivity, e.g. $K_{lat}/K_{el} > 10$. Therefore, there is potential to substantially improve the figure of merit of $Si_{0.8}Ge_{0.2}$ if one can find a way to reduce its lattice thermal conductivity.

## VII. Conclusion

The Lorenz number is the ratio of two thermodynamic quantities, i.e. the electronic contribution to the specific heat capacity to the concentration of the mobile charge carriers, or the electronic contribution to the specific heat capacity to the electrochemical capacitance of the electron system. As a result, the Lorenz number can be calculated directly from the electron density of states using DFT calculations, thereby providing a significant simplification over the conventional Boltzmann transport theory that requires the details of scattering mechanisms.

**Acknowledgements**

This work was supported by the Computational Materials Sciences Program funded by the US Department of Energy, Office of Science, Basic Energy Sciences, under Award Number DE-SC0020145 (Chen and Wang) and partially supported by National Science Foundation (NSF) through Grant No. CMMI-1825538 (Wang, Shang, and Liu). First-principles calculations were carried out partially on the resources of NERSC supported by the Office of Science of the US Department of Energy under contract No. DE-AC02-05CH11231, and partially on the resources of Extreme Science and Engineering Discovery Environment (XSEDE) supported by NSF with Grant No. ACI-1053575. We would like to thank professor Ravi of California Institute of Technology for critical reading of this manuscript.

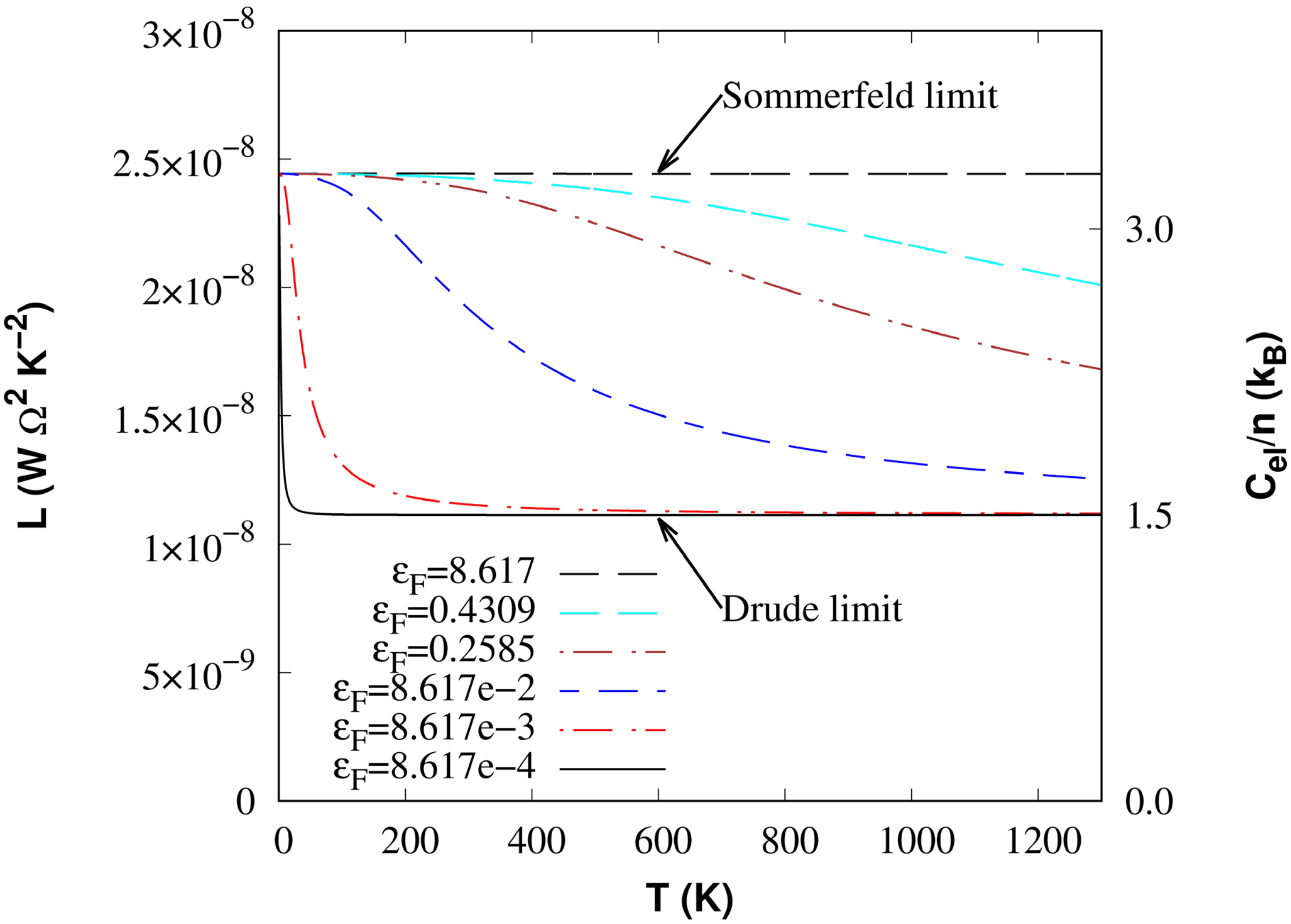

**Figure 1**. Calculated Lorenz numbers with the free electron gas model at $\boldsymbol{\varepsilon_F} = 8.617\times \mathbf{10^{-4}}$, $8.617\times \mathbf{10^{-3}}$, $8.617\times \mathbf{10^{-2}}$, 0.2585, 0.4309, and 8.617 eV.

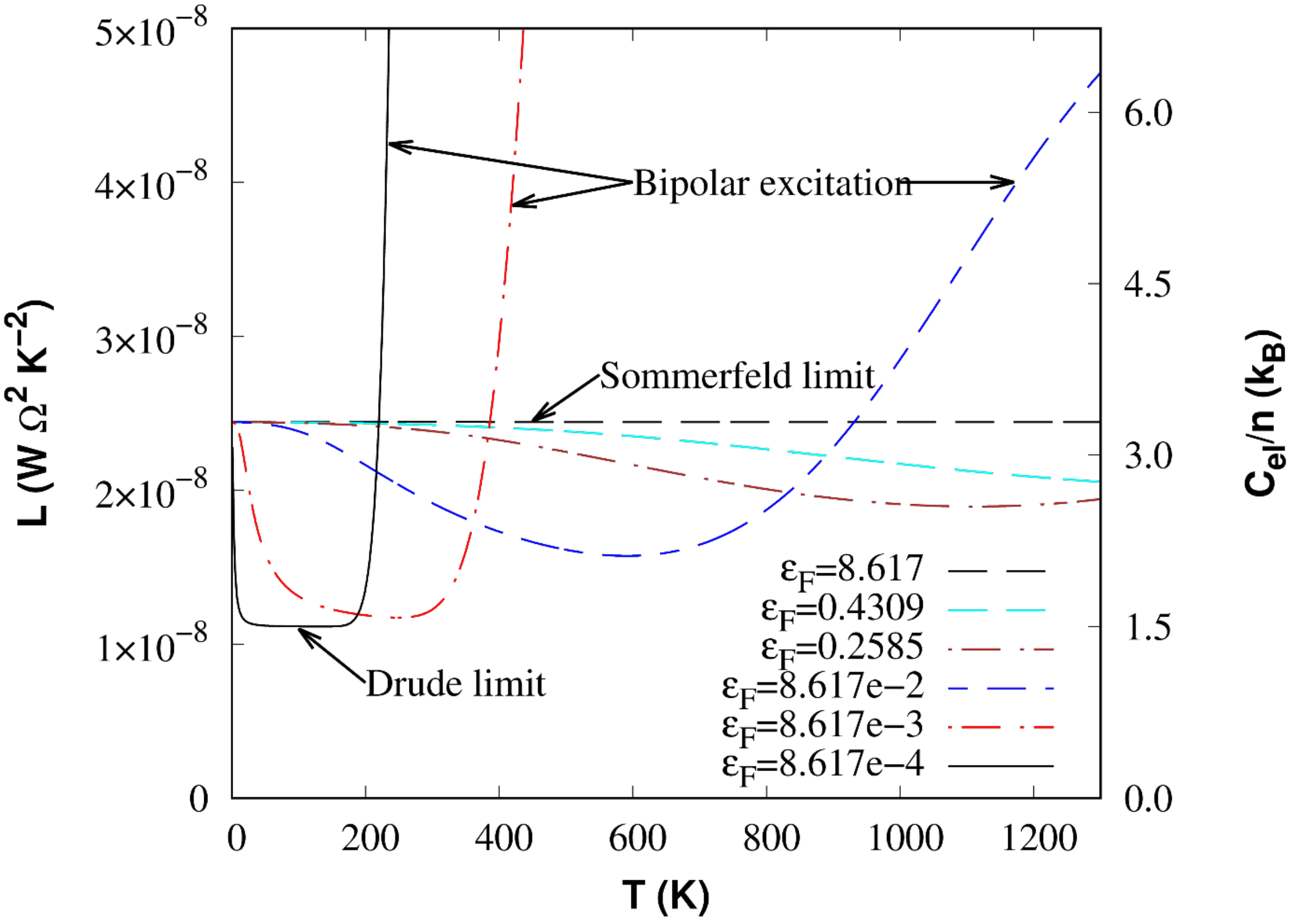

**Figure 2**. Calculated Lorenz numbers with the ideal semiconductor model at $\boldsymbol{\varepsilon_F} = 8.617\times \mathbf{10^{-4}}$, $8.617\times \mathbf{10^{-3}}$, $8.617\times \mathbf{10^{-2}}$, 0.2585, 0.4309, and 8.617 eV.

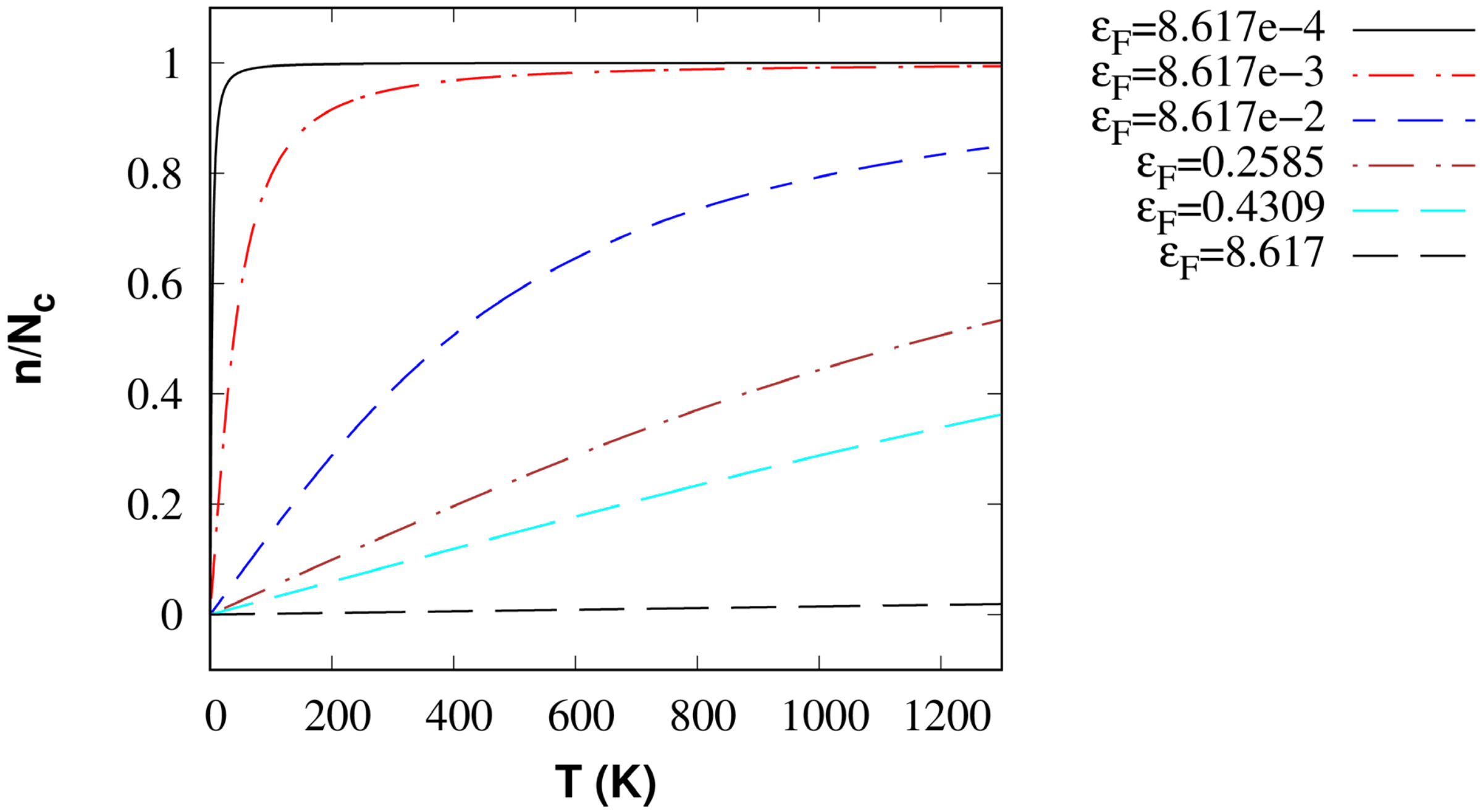

Figure 3. Normalized mobile electronic thermal carriers with the free electron gas model at $\boldsymbol{\varepsilon_F} =$ 8.617× $\mathbf{10^{-4}}$, 8.617× $\mathbf{10^{-3}}$, 8.617× $\mathbf{10^{-2}}$, 0.2585, 0.4309, and 8.617 eV.

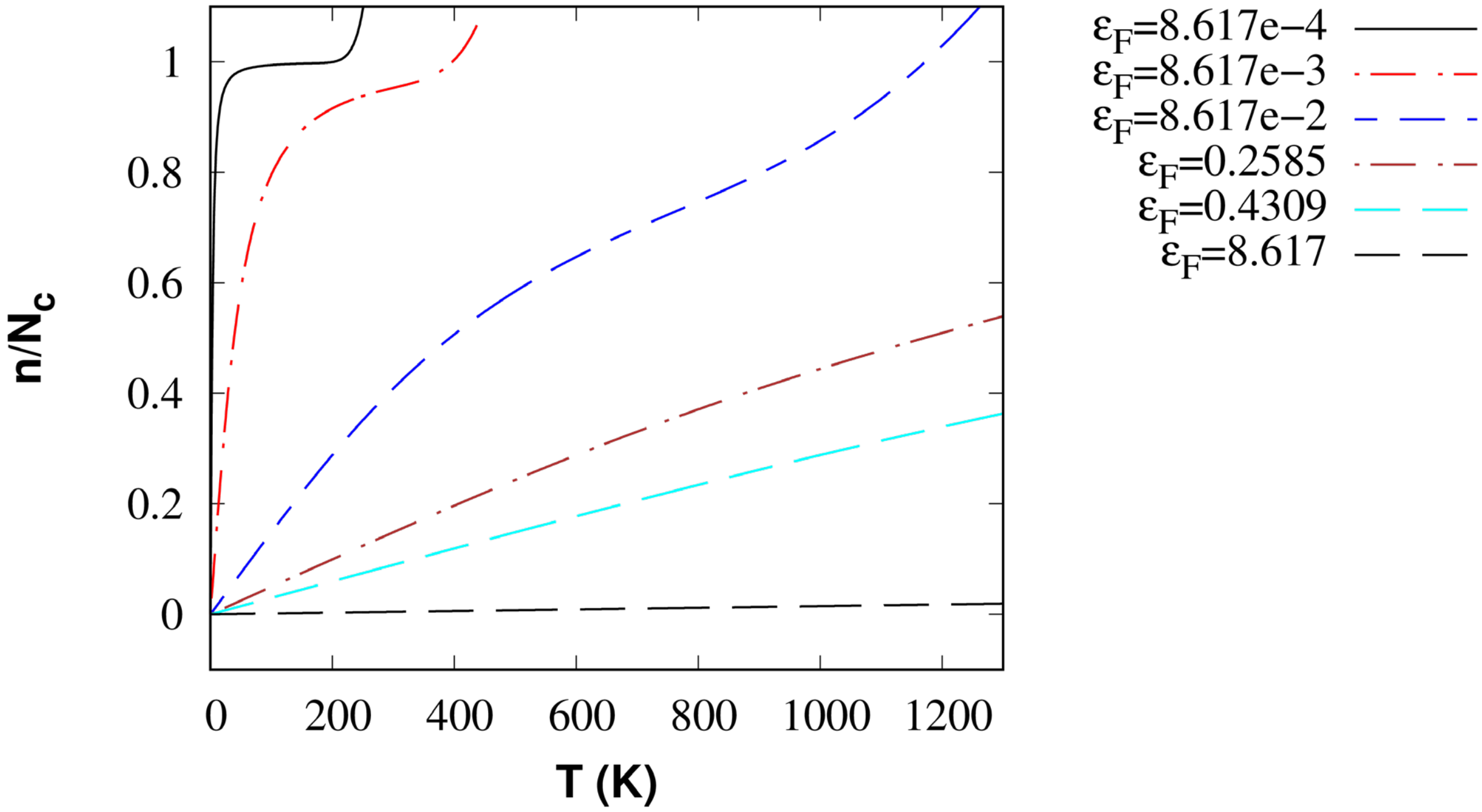

**Figure 4**. Normalized mobile electronic thermal carriers with a semiconductor model at $\boldsymbol{\varepsilon_F} =$ 8.617× $\mathbf{10^{-4}}$, 8.617× $\mathbf{10^{-3}}$, 8.617× $\mathbf{10^{-2}}$, 0.2585, 0.4309, and 8.617 eV.

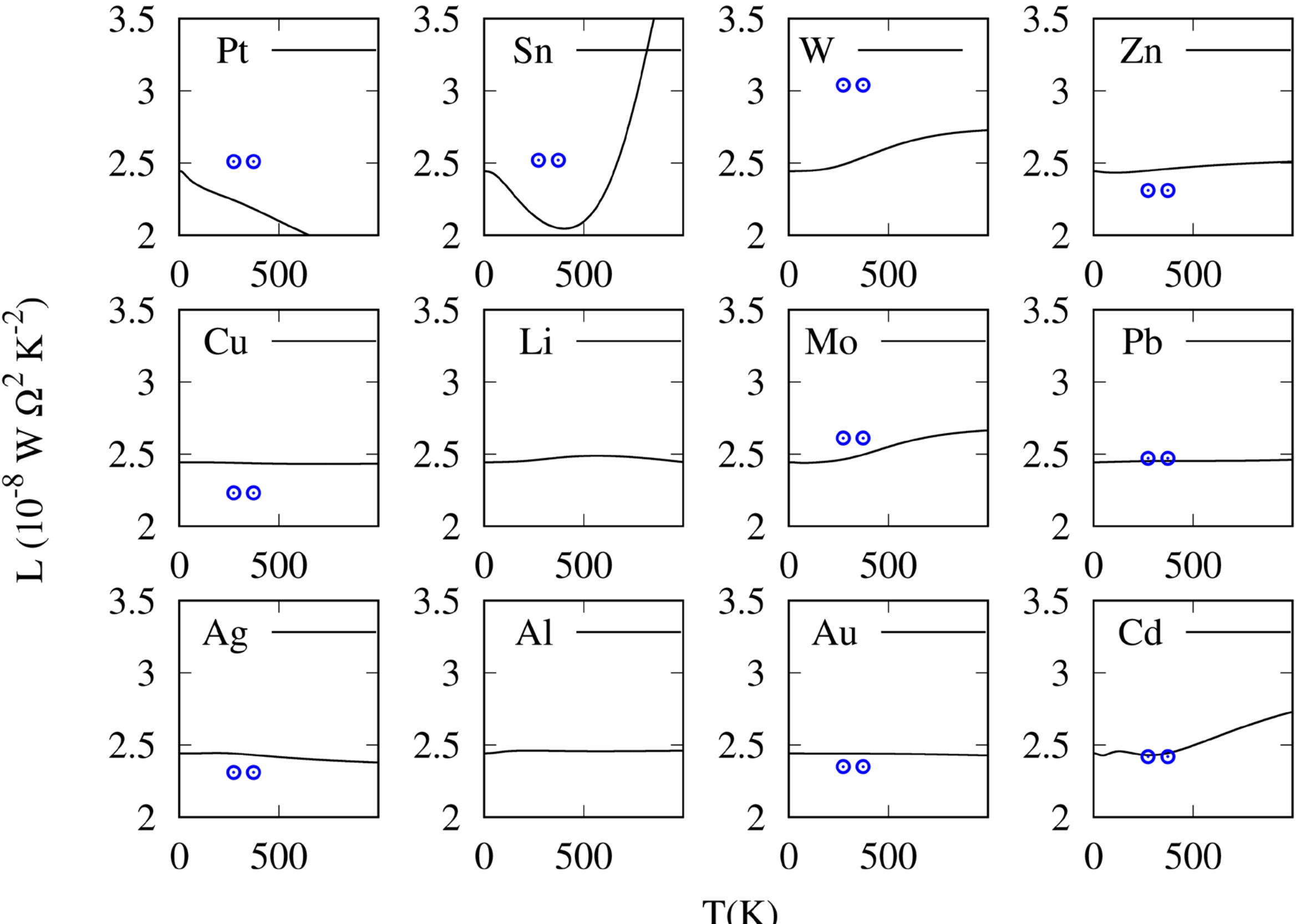

**Figure 5**. Calculated Lorenz numbers for pure Ag, Al, Au, Cd, Cu, Li, Mo, Pb, Pt, Sn, W, and Zn metals in their ground state crystal structures. The circles are experimental data from Kittel's textbook. [33]

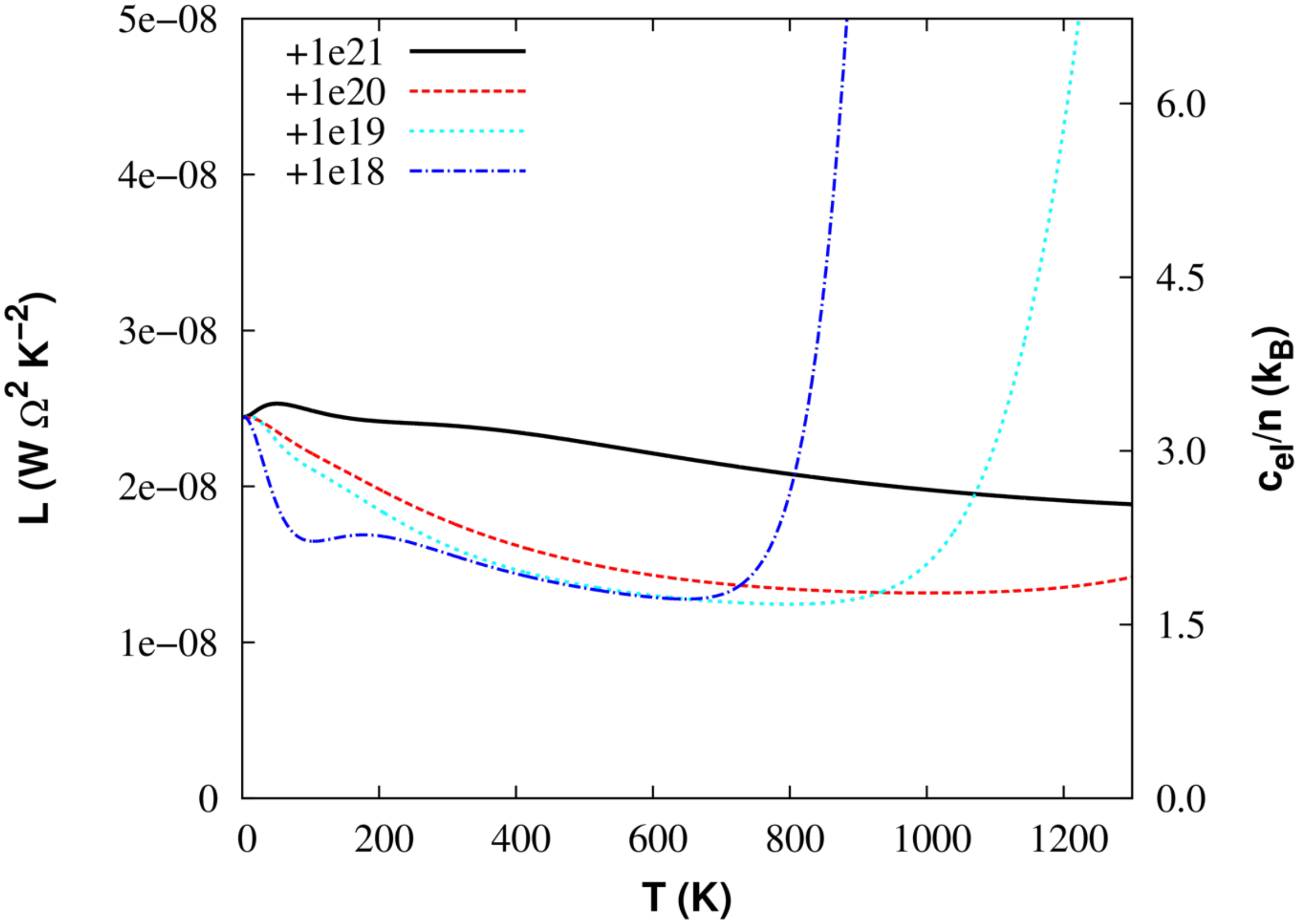

**Figure 6**. Calculated Lorenz numbers of $Si_{0.8}Ge_{0.2}$ at *p*-type doping levels of $1\times10^{18}$, $1\times10^{19}$, $1\times10^{20}$, and $1\times10^{21}$ *e*/cm$^3$, respectively.

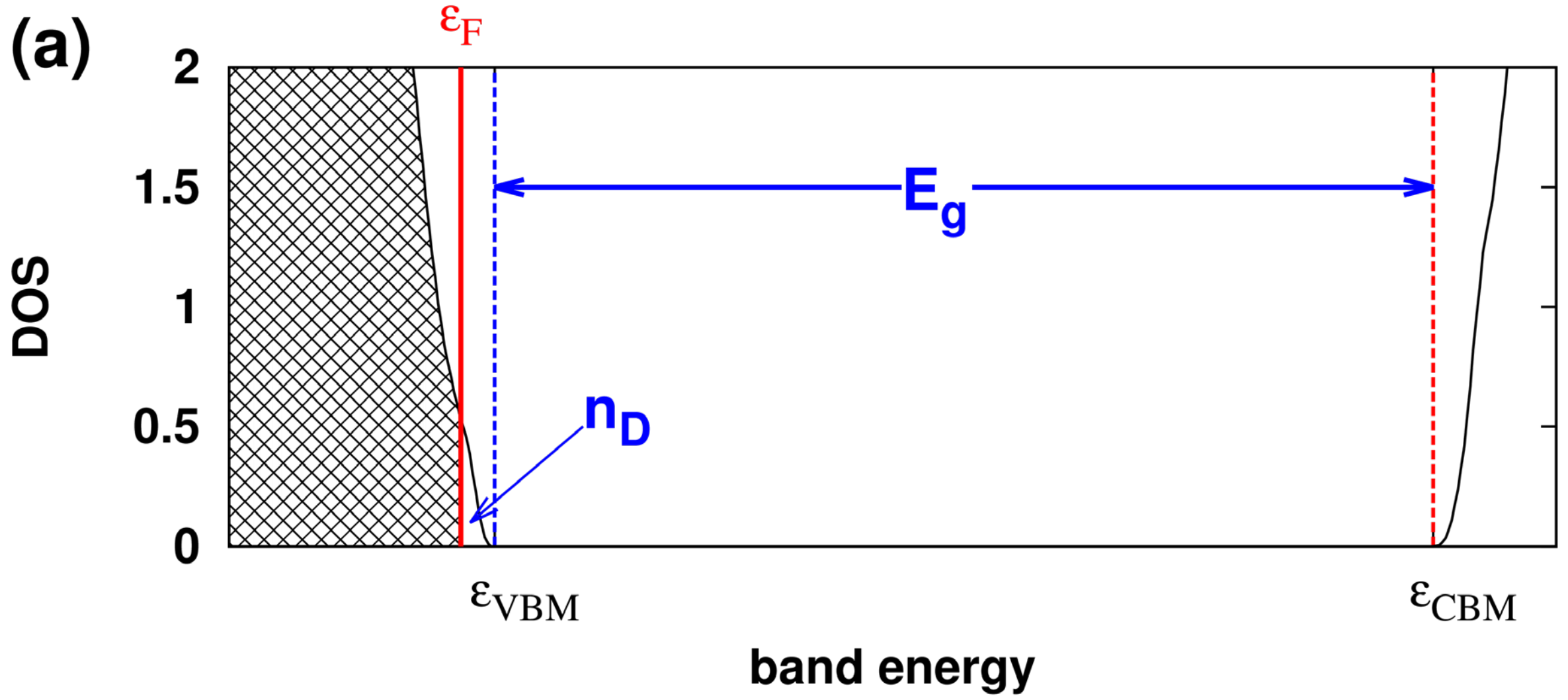

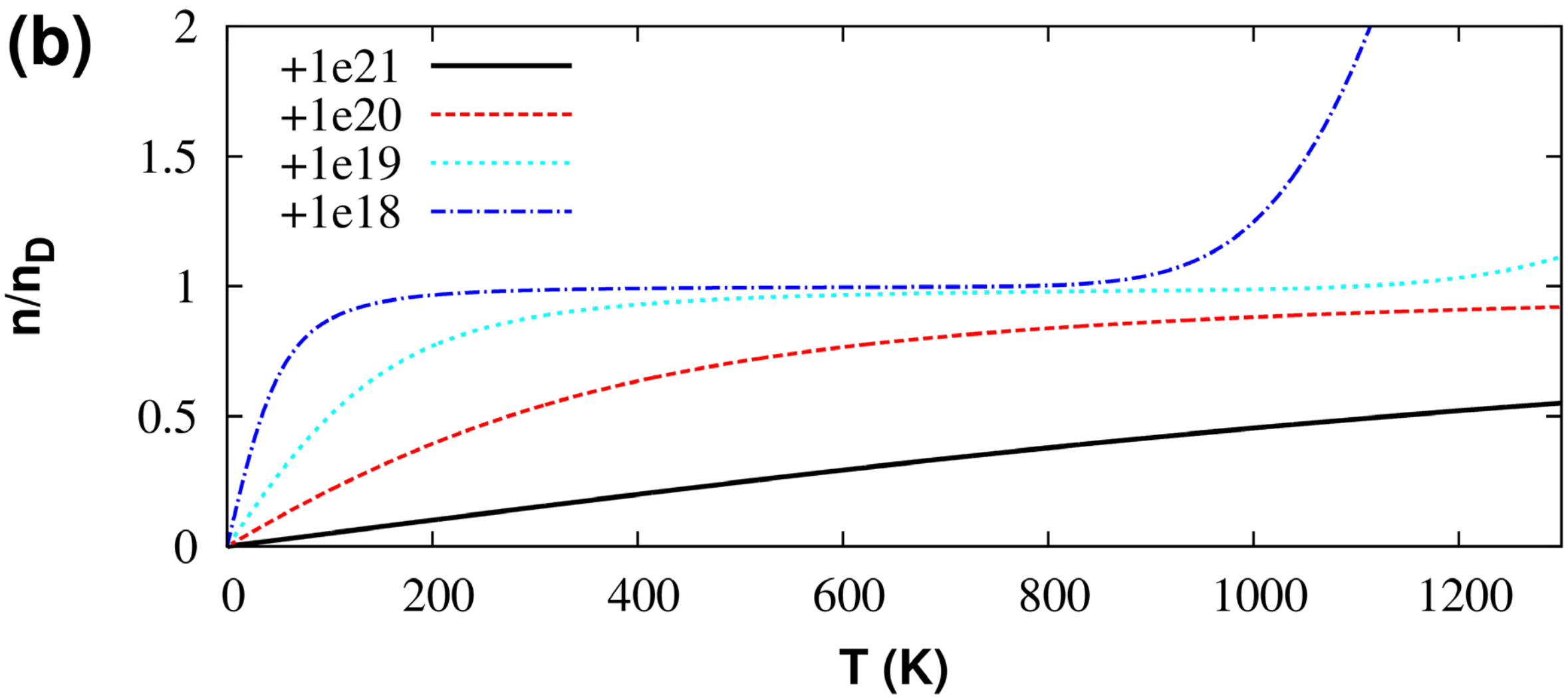

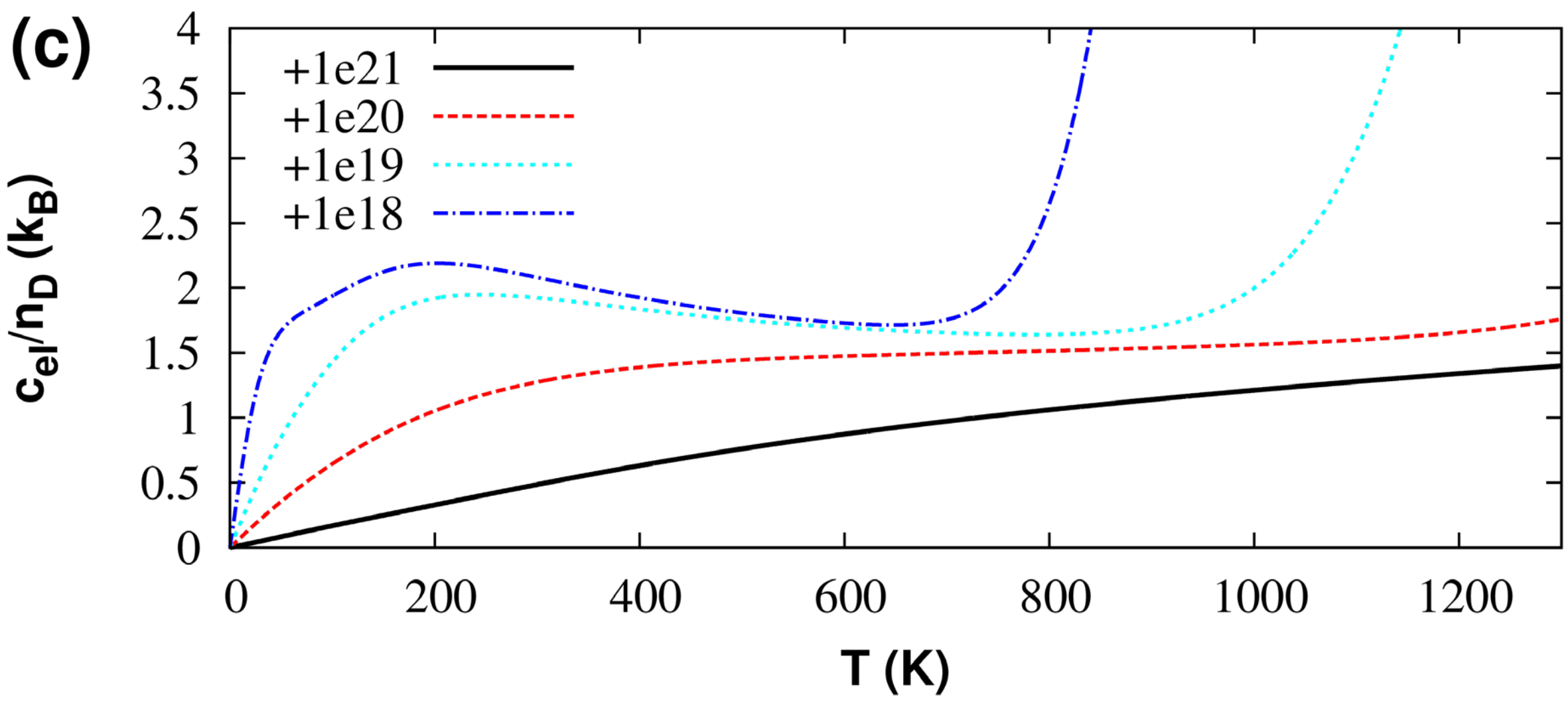

**Figure 7**. (a) Schematic illustration of doping using electron density of states where shadow area marks the occupied electronic states at 0 K, $\boldsymbol{\varepsilon_F}$ the Fermi energy, $\boldsymbol{\varepsilon_{VBM}}$ the valence band maximum, and $\boldsymbol{\varepsilon_{CBM}}$ the conduction band minimum; (b) Calculated $\boldsymbol{n/n_D}$, and(c) the calculated $\boldsymbol{c_{el}/n_D}$ at *p*-type doping levels of $1\times10^{18}$, $1\times10^{19}$, $1\times10^{20}$, and $1\times10^{21}$ $e$/cm$^3$, respectively.

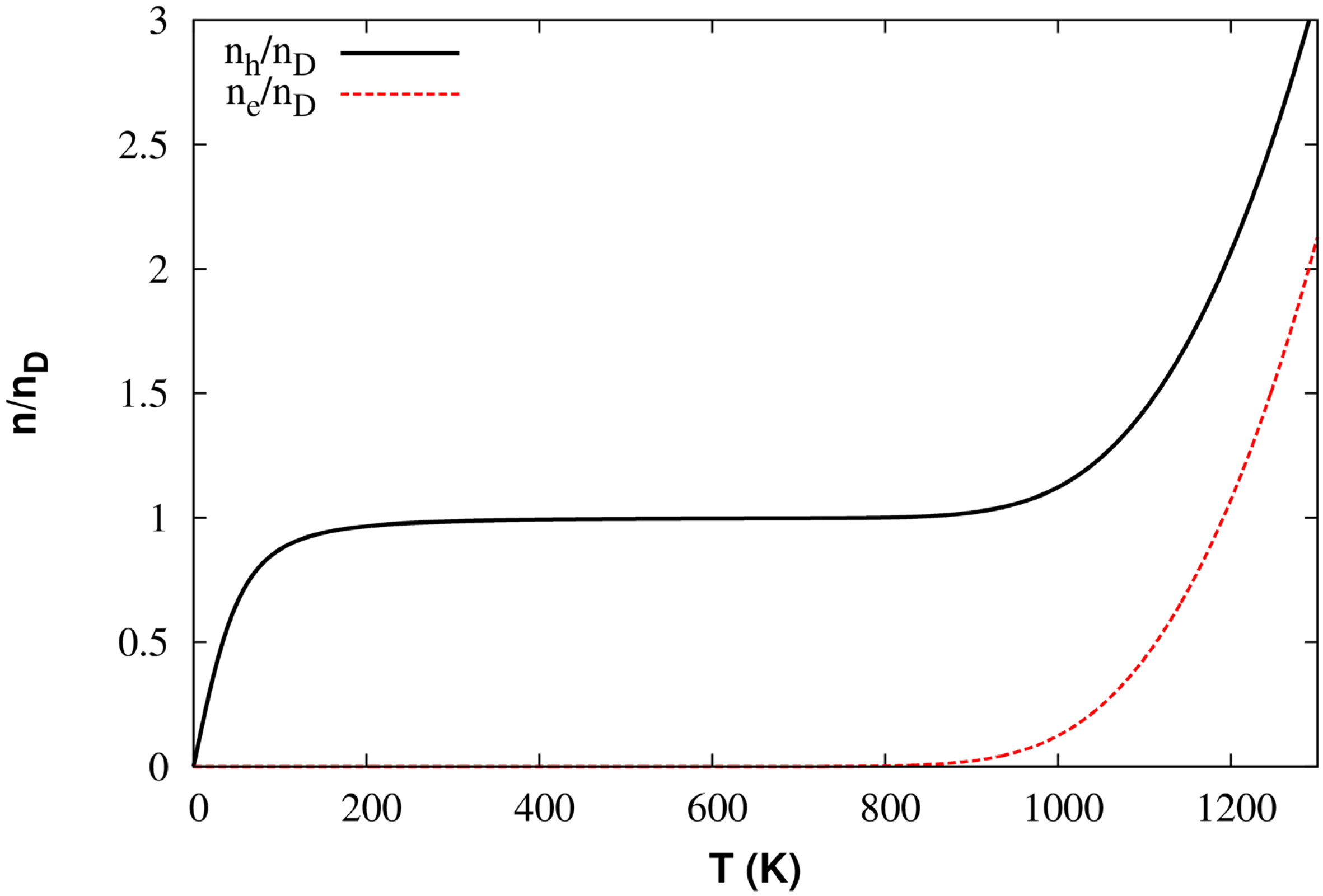

**Figure 8**. Temperature evolutions of $\boldsymbol{n_h/n_D}$ and $\boldsymbol{n_e/n_D}$ at *p*-type doping levels of $1\times10^{18}$ *e*/cm$^3$.

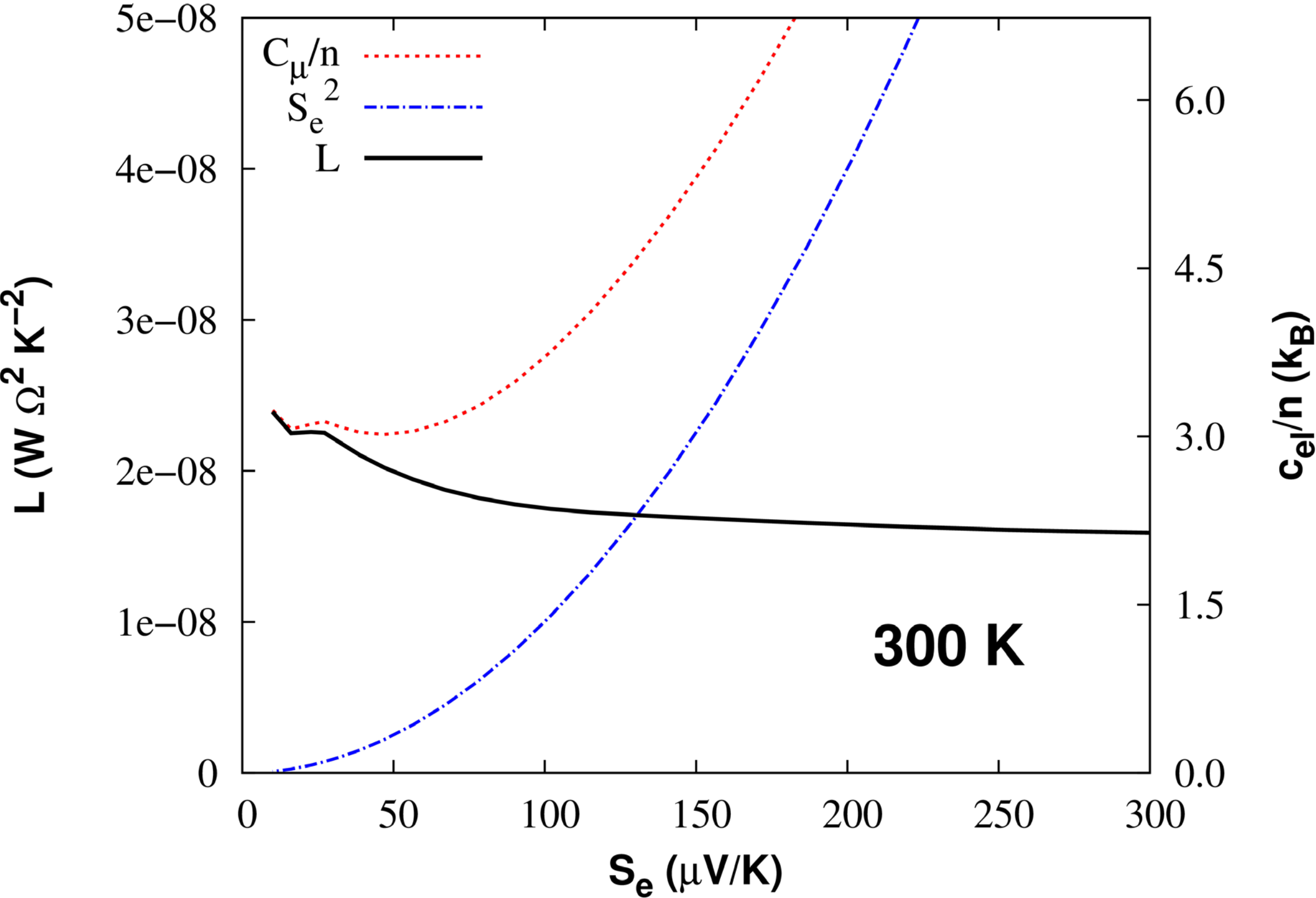

**Figure 9**. $C_\mu/n$ and $S_e^2$ contributions to the Lorenz numbers for *p*-type $Si_{0.8}Ge_{0.2}$ at 300 K.

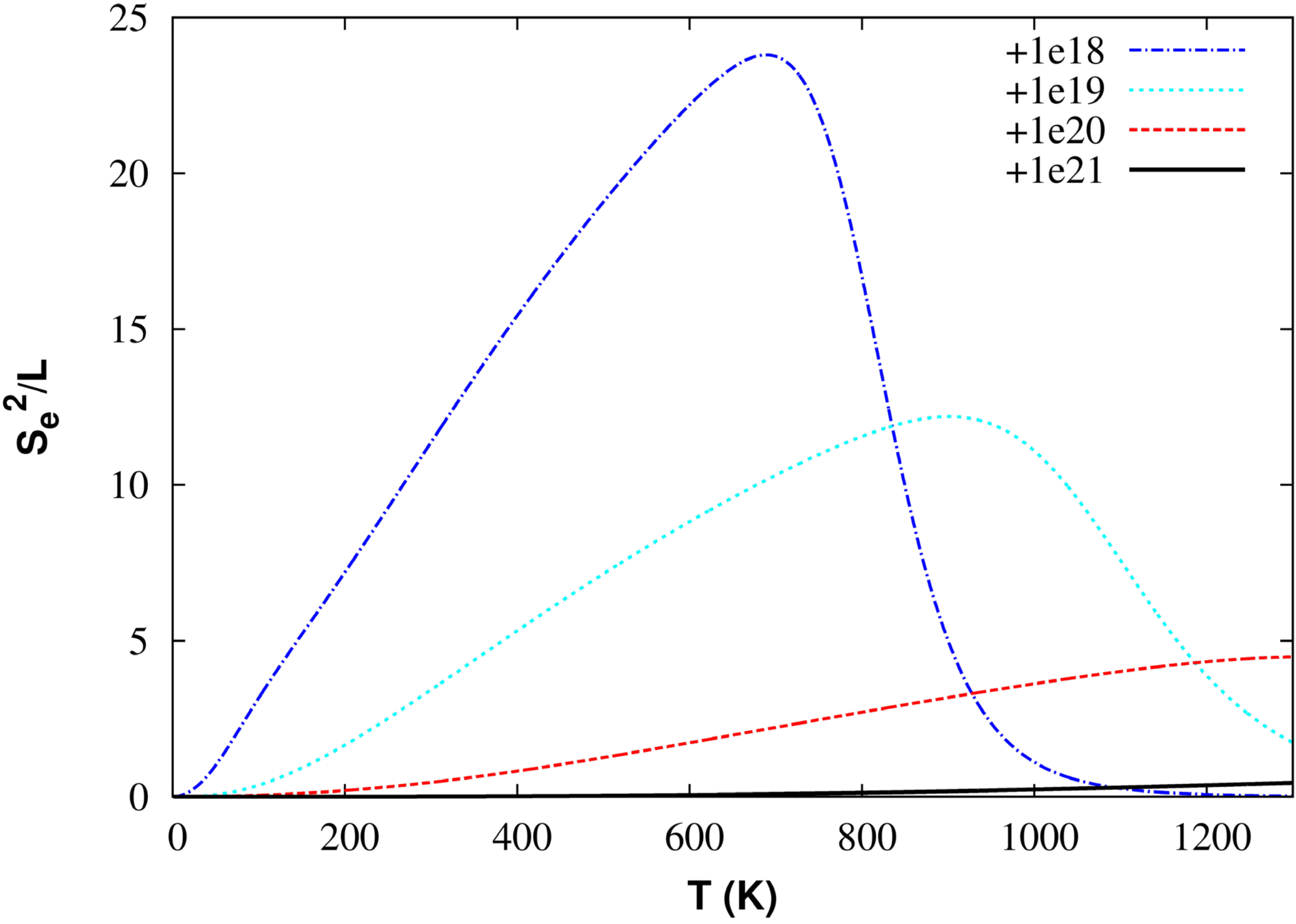

Figure 10. Calculated $zT_{el} = S_e^2/L$ at *p*-type doping levels of $1\times10^{18}$, $1\times10^{19}$, $1\times10^{20}$, and $1\times10^{21}$ $e$/cm$^3$, respectively.